\documentclass[aps,pra,reprint,superscriptaddress,longbibliography,floatfix]{revtex4-2}
\usepackage[utf8]{inputenc}
\usepackage{amsmath,amssymb,bm,graphicx,booktabs}
\usepackage{placeins}
\usepackage[colorlinks=true,linkcolor=blue,citecolor=blue,urlcolor=blue]{hyperref}
\newcommand{\Tr}{\operatorname{Tr}}
\newcommand{\sech}{\operatorname{sech}}
\newcommand{\Id}{\mathbb{I}}
\makeatletter
\AtBeginDocument{\immediate\write\@auxout{\string\citation{BibliographyControl}}}
\makeatother
\newcommand{\E}{\mathbb{E}}
\newcommand{\Var}{\operatorname{Var}}
\newcommand{\diag}{\operatorname{diag}}
\hypersetup{pdftitle={Ordering controlled by the measurement-feedback interval in quantum Boltzmann samplers}}
\begin{document}
\title{Ordering controlled by the measurement-feedback interval in quantum Boltzmann samplers}
\author{Peng Wang}
\email{wp002005@163.com}
\affiliation{School of Computer and Artificial Intelligence, Southwest Minzu University, Chengdu 610225, China}
\affiliation{Centre for Quantum Technologies, National University of Singapore, Singapore 117543, Singapore}

\author{Yu-Xuan Zhang}
\affiliation{School of Physics, Nankai University, Tianjin 300071, China}
\affiliation{Centre for Quantum Technologies, National University of Singapore, Singapore 117543, Singapore}

\author{Hai-Tao Ding}
\affiliation{Centre for Quantum Technologies, National University of Singapore, Singapore 117543, Singapore}
\affiliation{MajuLab, CNRS-UNS-NUS-NTU International Joint Research Unit, Singapore UMI 3654, Singapore}

\author{Leong-Chuan Kwek}
\email{kwekleongchuan@nus.edu.sg}
\affiliation{Centre for Quantum Technologies, National University of Singapore, Singapore 117543, Singapore}
\affiliation{MajuLab, CNRS-UNS-NUS-NTU International Joint Research Unit, Singapore UMI 3654, Singapore}
\affiliation{National Institute of Education, Nanyang Technological University, Singapore 637616, Singapore}

\begin{abstract}
We show that the measurement--feedback interval alone can induce ordering in a quantum Boltzmann sampler with fixed programmed couplings and a fixed local dissipative rule. At zero environmental dephasing, repeated projective measurements remove transverse coherence regenerated during relaxation, thereby enhancing the local longitudinal response. We derive the exact finite-interval Markov kernel for successive measurement records and obtain the ordering threshold in the Curie--Weiss limit. For couplings below the equilibrium Curie--Weiss threshold, shortening the readout interval drives a continuous transition to an ordered stationary state even though the programmed Boltzmann distribution remains disordered. The stationary response and equation of state map exactly onto those of a continuously interacting model with dephasing, while the linear relaxation rates remain distinct at matched static response. Exact finite-size stationary calculations reveal the transition through bimodal magnetization distributions and mean-field critical scaling.
\end{abstract}

\maketitle
\section{Introduction}

The weights of a Boltzmann machine specify a target probability law, but reproducing that law depends on the sampling dynamics~\cite{Ackley1985,Glauber1963}. Quantum approaches seek to realize such distributions through state encoding, thermal-state preparation, or programmable hardware~\cite{Somma2007,Amin2018,Benedetti2017,Temme2011,Wild2021,Wild2021PRA}, while engineered dissipation provides a complementary route in open quantum systems~\cite{Kraus2008,Diehl2008,Verstraete2009}, including neural and associative-memory dynamics~\cite{Rotondo2018,Fiorelli2019}. When measurement outcomes determine subsequent inputs, measurement and feedback themselves become part of the sampling dynamics, so their timing can alter the sampled distribution relative to the programmed law.

Measurement backaction and feedback can alter collective behavior,
generating magnetic correlations~\cite{Mazzucchi2016},
inducing quantum and magnetic transitions~\cite{Ivanov2020,Hurst2020},
and producing phase transitions and symmetry breaking in adaptive
monitored systems~\cite{Friedman2023,Ravindranath2023,Iadecola2023,Hauser2024}.
Periodic resetting measurements can drive ordering transitions in
dissipative Floquet systems~\cite{Sierant2022}, while
measurement--feedback-driven transitions have also been observed on
a superconducting processor~\cite{Wu2026}.
For sampling networks, the update dynamics also matters:
discrete-time measurement feedback affects the sampled configurations
in coherent Ising machines~\cite{Yamamura2017,Ng2022}.
In the partial-reset qubit model underlying the present study,
we previously showed that environmental dephasing enhances the
local response and shifts the ordering threshold of a continuously
interacting network~\cite{Zhang2026}.

Periodic measurement and feedback differ from continuous dephasing
because each readout also updates the fields governing the subsequent
evolution. We therefore ask how the measurement--feedback interval
controls ordering and relaxation at fixed programmed weights and
local dissipative parameters. We retain the partial-reset channel
of our previous work and set the environmental dephasing rate to zero.
All qubits are measured simultaneously in $Z$ at intervals $\Delta$,
and the recorded outcomes set classical feedback fields that remain
fixed until the next readout. Varying $\Delta$ thus changes both the
readout interval and the duration for which the feedback fields are held.

We derive an exact finite-interval transition kernel for the
measurement records. In the Curie--Weiss limit, we identify a range
of couplings below the equilibrium threshold for which shortening
the readout interval drives a continuous bifurcation to two stable
magnetized branches, while the programmed Boltzmann law remains
disordered. The local stationary response and Curie--Weiss equation
of state map onto those of the continuous-dephasing model, but the
linear relaxation rates differ at matched static response.
Exact finite-size stationary calculations show the corresponding
bimodal magnetization distribution without postselection, and the
kernel determines the protocol-dependent amplitude of critical
fluctuations. A separate comparison with continuous classical
updates identifies the effect of holding feedback fields fixed
on multispin transition probabilities.

Section~II derives the local response and the record kernel. Section~III establishes the bifurcation and stability of the Curie--Weiss limiting map, including a relaxation-rate comparison at matched static response. Section~IV computes finite-size stationary statistics and their critical scaling. Section~V gives the linear instability threshold for dense population networks.

\section{Measured feedback dynamics}
\subsection{Programmed law and physical protocol}
The programmed reference law is
\begin{equation}
 p_{\rm tar}(\bm z)\propto
 \exp\!\left(\sum_i b_i z_i+\sum_{i<j}J_{ij}z_i z_j\right),
 \qquad z_i=\pm1 .
 \label{eq:target}
\end{equation}
All fields and couplings are dimensionless. At times $k\Delta$, every qubit is measured in $Z$, with $Z|0\rangle=|0\rangle$. The outcomes set classical fields $a_i=b_i+\sum_{j\ne i}J_{ij}z_j$, which are held fixed during the next interval. Each qubit then evolves independently with $\mathcal L_a=\nu(\Phi_a-\Id)$, a completely positive Markov generator~\cite{Gorini1976,Lindblad1976}. The local channel is inherited from Ref.~\cite{Zhang2026}:
\begin{align}
 |\psi_a\rangle&=\sqrt{\frac{1+u}{2}}|0\rangle+
                  \sqrt{\frac{1-u}{2}}|1\rangle,\quad u=\tanh a,\nonumber\\
 A_0&=P+sQ,\qquad
 A_1=\sqrt{1-s^2}|\psi_a\rangle\langle\psi_a^\perp|,
 \label{eq:channel}
\end{align}
where $P=|\psi_a\rangle\langle\psi_a|$, $Q=\Id-P$, $\Phi_a(\rho)=\sum_\mu A_\mu\rho A_\mu^\dagger$, and $0\le s<1$. The environmental dephasing rate is $\gamma=0$ throughout. Measurements are ideal, simultaneous, and instantaneous, feedback has no dead time, and independent local environments implement the channel updates at rate $\nu>0$.

Changing $\Delta$ changes both the readout times and the duration for which each feedback field is held. The channel family $\mathcal L_a$, its parameters $s,\nu$, and the programmed weights remain fixed; the resulting network evolution depends on the schedule. Here ``Boltzmann sampler'' denotes a network programmed from Eq.~\eqref{eq:target}. Its synchronous updates generally do not preserve that reference law, even at $s=0$~\cite{Peretto1984,Grinstein1985}.

After each full readout, the conditional qubit state is a product of $Z$ eigenstates. Independent local evolution preserves this product structure until the next readout. Averaging over records therefore gives a separable qubit state. Local coherence affects the transition probabilities, while correlations between sites are carried by the classical feedback record.

\subsection{Exact finite-interval kernel}

At fixed input, let $q=\sech a$ and $\bm n=(q,0,u)$, the unit Bloch vector of $|\psi_a\rangle$. The channel in Eq.~\eqref{eq:channel} acts on a Bloch vector as
\begin{equation}
 \bm v\longmapsto s\bm v-s(1-s)(\bm n\cdot\bm v)\bm n+(1-s^2)\bm n.
 \label{eq:blochchannel}
\end{equation}
Completeness follows from $A_0^\dagger A_0=P+s^2Q$ and $A_1^\dagger A_1=(1-s^2)Q$. The generator is in Lindblad form with jump operators $\sqrt\nu A_\mu$~\cite{Gorini1976,Lindblad1976}. The transverse and longitudinal relaxation rates relative to $\bm n$ are $\kappa=\nu(1-s)$ and $\kappa(1+s)$, respectively. Thus
\begin{equation}
 \dot{\bm v}=\kappa[(1+s)\bm n-(\Id+s\bm n\bm n^{\mathsf T})\bm v].
\end{equation}
Since $\bm n$ is an eigenvector of the drift matrix, the finite-time solution is
\begin{equation}
\begin{aligned}
 \bm v(\Delta)&=\bm n+R_\Delta[\bm v(0)-\bm n],\\
 R_\Delta&=E\Id+(L-E)\bm n\bm n^{\mathsf T},\\
 E&=e^{-\kappa\Delta},\qquad L=e^{-\kappa(1+s)\Delta}.
\end{aligned}
\label{eq:propagator}
\end{equation}
Inserting $\bm v(0)=(0,0,z)$ gives
\begin{equation}
 \E[z'|z,a]=[E+(L-E)u^2]z+(1-L)u=U(a)z+D(a).
 \label{eq:conditionalmean}
\end{equation}
The exact binary transition kernel is
\begin{equation}
 p_\Delta(z'|z,a)=\frac{1+z'[U(a)z+D(a)]}{2},
 \label{eq:kernel}
\end{equation}
where $U(a)=E+(L-E)\tanh^2a$ and $D(a)=(1-L)\tanh a$. For the network this gives
\begin{equation}
 P_\Delta(\bm z'|\bm z)=\prod_i p_\Delta(z_i'|z_i,a_i(\bm z)).
 \label{eq:networkkernel}
\end{equation}
All dependence between sites enters through the previous record and the resulting fields. Complete positivity ensures that these numbers lie in $[0,1]$. They are strictly inside this interval for finite $a$, $\Delta>0$, and $s<1$.

\subsection{Response enhancement under repeated readout}
For a clamped field the invariant polarization is
\begin{equation}
 \bar z_\Delta(a)=\frac{D(a)}{1-U(a)}
 =\frac{(1+\alpha_\Delta)\tanh a}{1+\alpha_\Delta\tanh^2a},
 \quad \alpha_\Delta=\frac{E-L}{1-E}.
 \label{eq:measuredactivation}
\end{equation}
This is the invariant polarization of the repeated measured chain. The unmeasured qubit's stationary polarization is instead $\tanh a$ at zero dephasing.

Let $x=\kappa\Delta>0$. Then $\alpha_\Delta=(1-e^{-sx})/(e^x-1)$. For $s>0$,
\begin{equation}
 \frac{d\log\alpha_\Delta}{dx}=
 \frac{s}{e^{sx}-1}-\frac{1}{1-e^{-x}}<0,
\end{equation}
because the first term is below $1/x$ and the second is above $1/x$. The limits are $\alpha_{0^+}=s$ and $\alpha_\infty=0$. At $s=0$, $\alpha_\Delta=0$ for every interval. The excluded endpoint $s=1$ makes the channel the identity and destroys relaxation.

To see the microscopic origin of this change, write $v_x$ and $v_z$ for the two relevant Bloch components during an interval. The longitudinal equation in the measurement basis is
\begin{equation}
 \dot v_z=\kappa[(1+s)u-(1+su^2)v_z-squv_x].
 \label{eq:coherencefeedback}
\end{equation}

\begin{figure}[tb]
    \centering
    \includegraphics[width=\columnwidth]{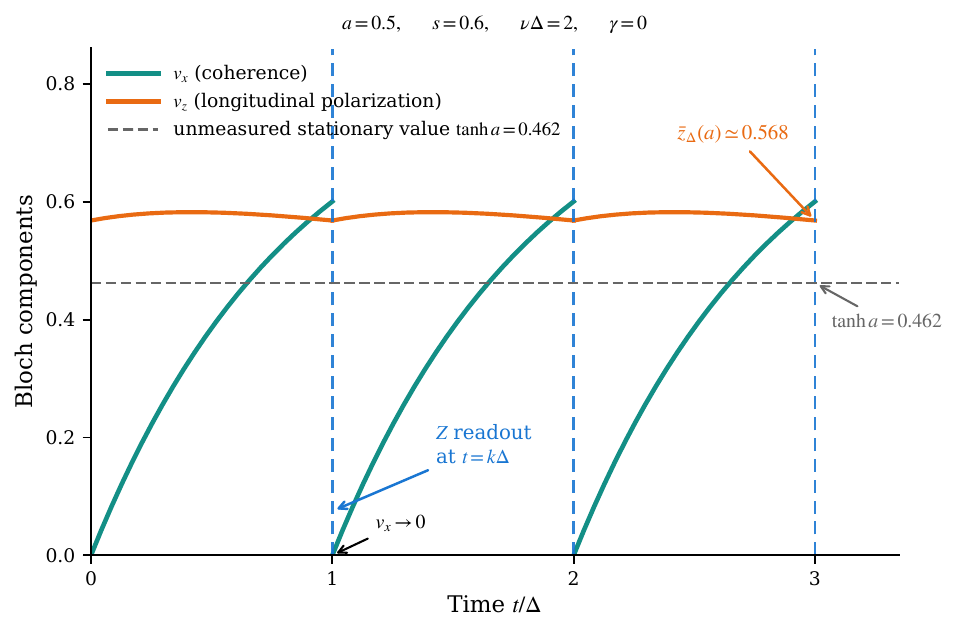}
    \caption{
    Local dynamics underlying the response enhancement under
    periodic readout.
    Outcome-averaged Bloch components in the periodic steady
    state for a clamped input $a=0.5$, with $s=0.6$,
    $\nu\Delta=2$, and zero environmental dephasing. The outcome-averaged dynamics follow Eq. ~\eqref{eq:propagator} initialized immediately after a readout at $v_x=0$ and $v_z=\bar z_\Delta(a)$.
    Blue dashed lines mark projective $Z$ readouts, which
    reset $v_x$ to zero while leaving the outcome-averaged
    $v_z$ unchanged.
    Between readouts, transverse coherence rebuilds and
    $v_z$ evolves continuously.
    Orange dots mark the stationary readout polarization
    $\bar z_\Delta(a)\simeq0.568$, given by Eq.~\eqref{eq:measuredactivation};
    this is not a time average over the interval.
    The horizontal gray dashed line marks the unmeasured
    stationary polarization, $\tanh a\simeq0.462$.
    }
    \label{fig:local}
\end{figure}

For the unmeasured stationary qubit, $v_x=q$ and $v_z=u$; the last term participates in the cancellation that fixes $v_z=\tanh a$. A projective $Z$ measurement removes $v_x$ and prepares $v_z=z$. The transverse component subsequently rebuilds, but the next measurement removes it again. For positive input and a positive transverse component, this removal suppresses the negative contribution $-\kappa squv_x$ to the polarization drift. Figure~\ref{fig:local} illustrates this mechanism for a clamped input $a=0.5$, with
$s=0.6$ and $\nu\Delta=2$. The outcome-averaged transverse component is
reset to zero at each readout and subsequently rebuilds, whereas the
longitudinal component returns to the stationary readout value
$\bar{z}_{\Delta}(a)\simeq 0.568$. This exceeds the unmeasured stationary
polarization $\tanh a\simeq 0.462$, directly visualizing the response
enhancement produced by repeated readout.

The effect is especially transparent in linear response. At $a=0$, the average measured polarization decays by the factor $E$ per round, whereas a small input generates a polarization increment $(1-L)a$. The two factors arise from distinct decay channels of the same local quantum map. The stationary gain is therefore
\begin{equation}
 G_\Delta=\left.\frac{d\bar z_\Delta}{da}\right|_{a=0}
 =\frac{1-L}{1-E}=1+\alpha_\Delta.
 \label{eq:gain}
\end{equation}
For $s>0$ and finite $\Delta$, $L<E$ and $G_\Delta>1$. At zero clamped input the stationary record remains unbiased. Collective order arises when feedback converts a small measured magnetization into a reinforcing input.

\subsection{Relation to the dephasing-driven model: static correspondence and finite-time dynamics}\label{sec:comparison}

For comparison, the dephasing-controlled implementation built from the same local channel has the clamped stationary polarization reported in Ref.~\cite{Zhang2026}, with
\begin{equation}
 \alpha_r=\frac{sr}{1+r},
\end{equation}
where $r=\gamma/[\nu(1-s^2)]$ is the dephasing strength used in that model. For $0<s<1$ and finite $\Delta$, $0<\alpha_\Delta<s$, so
\begin{equation}
 r_{\rm eff}(\Delta)=\frac{\alpha_\Delta}{s-\alpha_\Delta}
 \quad\Longrightarrow\quad
 \alpha_{r_{\rm eff}}=\alpha_\Delta.
 \label{eq:staticmapping}
\end{equation}
This mapping matches the full clamped response. It also matches the Curie--Weiss equation of state in Eq.~\eqref{eq:eos}, including the critical coupling and nonzero magnetization branches. The common static equation does not fix the time evolution.

\paragraph*{Selective readout and stored feedback.}
For a projective $Z$ readout with projectors $\Pi_z$, the selective operation associated with outcome $z$ is
\begin{equation}
 \mathcal I_z(\rho)=\Pi_z\rho\Pi_z,
 \qquad p_z=\Tr[\mathcal I_z(\rho)].
 \label{eq:instrument}
\end{equation}
If the outcome is ignored, summing over $z$ gives the nonselective map
\begin{equation}
 \mathcal M_Z(\rho)=\sum_z\Pi_z\rho\Pi_z,
 \label{eq:nonselective}
\end{equation}
which removes coherence in the readout basis. In our protocol the controller retains the outcome to set the next field. Averaging over records after applying that feedback preserves the outcome-dependent evolution; omitting the feedback would define a different process. Between readouts, transverse coherence rebuilds under $\mathcal L_a$, whereas continuous dephasing damps it throughout the interval.

In the continuously interacting comparison model, let $p_N(\bm z)=\langle\bm z|\rho_N|\bm z\rangle$, let $\bm z^i$ denote configuration $\bm z$ with spin $i$ reversed, and write $u_i=\tanh a_i$, $q_i=\sech a_i$. In the dimensionless time $\tau=\kappa t$, its population equation takes the form
\begin{equation}
\begin{aligned}
 \partial_\tau p_N(\bm z)=\sum_i\{&w_i(\bm z^i)p_N(\bm z^i)-w_i(\bm z)p_N(\bm z)\\
 &-s z_i u_i q_i\,\mathrm{Re}\langle\bm z|\rho_N|\bm z^i\rangle\},
\end{aligned}
\label{eq:ref24population}
\end{equation}
with
\begin{equation}
 w_i(\bm z)=\frac12[1+s u_i(\bm z)^2-(1+s)z_i u_i(\bm z)].
 \label{eq:ref24rate}
\end{equation}
The explicit coherence term means that populations alone do not determine their subsequent evolution. By contrast, projective readout closes the present measurement record into the exact synchronous kernel in Eq.~\eqref{eq:networkkernel}, with fields fixed by the preceding record.

The rapid-readout limit gives a separate comparison with continuous classical updates. Expanding the one-round kernel at small $\Delta$ gives the continuous classical flip rate
\begin{equation}
 w(z\to-z\,|a)=\frac{\kappa}{2}
 [1+s\tanh^2a-z(1+s)\tanh a].
 \label{eq:rapidrate}
\end{equation}
This is also the physical single-spin rate in the strong-dephasing limit of the comparison model. The following test compares synchronous frozen-field evolution with the continuous classical process at these rates; it is not a calculation of the finite-dephasing model at $r_{\rm eff}(\Delta)$. Let $w_i(\bm z)$ denote Eq.~\eqref{eq:rapidrate} at the field produced by configuration $\bm z$, and let $\bm z^{ij}$ have distinct spins $i$ and $j$ reversed. The synchronous protocol gives
\begin{equation}
 P_\Delta(\bm z^{ij}|\bm z)
 =w_i(\bm z)w_j(\bm z)\Delta^2+O(\Delta^3),
 \label{eq:synchronousdouble}
\end{equation}
where both sites evolve using the fields stored from the initial record. A continuous-time single-spin process with generator $W$ and the same instantaneous rates instead gives
\begin{equation}
\begin{aligned}
 \bigl[e^{\Delta W}\bigr]_{\bm z^{ij},\bm z}
 =\frac{\Delta^2}{2}\{&w_i(\bm z)w_j(\bm z^i)\\
 &+w_j(\bm z)w_i(\bm z^j)\}+O(\Delta^3).
\end{aligned}
\label{eq:continuousdouble}
\end{equation}
For interacting sites the first flip changes the field entering the second rate. The two expressions therefore differ generically at order $\Delta^2$. This difference is present even at $s=0$, when the response enhancement vanishes: for two spins with $J_{12}=J>0$, starting from $++$, the coefficients of $\Delta^2$ are $\kappa^2(1-u)^2/4$ for the synchronous update and $\kappa^2(1-u^2)/4$ for the continuous process, where $u=\tanh J$. The test identifies the effect of the feedback schedule, not a uniquely quantum signature.

Equation~\eqref{eq:rapidrate} also shows that arbitrarily frequent readout leaves finite stochastic population dynamics. This is consistent with quantum Zeno dynamics for general operations~\cite{Misra1977,Facchi2002,Burgarth2020}. A comparison with the finite-dephasing model at matched response is given by the linear relaxation rates in Sec.~\ref{sec:ordering}.

\section{Measurement-induced ordering}\label{sec:ordering}

\subsection{Controlled thermodynamic limit}

Set $b_i=0$, $K>0$, $M_N=N^{-1}\sum_i z_i$, and $J_{ij}=K/N$ for $i\ne j$. Given a configuration, the local fields are $a_i=K(M_N-z_i/N)$. Conditional on that configuration, the next measured spins are independent. Since $U$ and $D$ are smooth with uniformly bounded derivatives on the accessible field interval,
\begin{align}
 \E[M_N'|\bm z]&=U(KM_N)M_N+D(KM_N)+O(N^{-1}),\nonumber\\
 \Var(M_N'|\bm z)&\le N^{-1}.
 \label{eq:lln}
\end{align}
The constants are uniform in the configuration at fixed $K,s,\Delta$. If $M_N(0)$ converges in probability to $m_0$, induction and conditional Chebyshev bounds imply convergence for every fixed number of rounds to
\begin{equation}
\begin{aligned}
 m_{k+1}&=f(m_k),\\
 f(m)&=Em+(1-L)\tanh(Km)\\
 &\quad -(E-L)m\tanh^2(Km).
\end{aligned}
\label{eq:mfmap}
\end{equation}
This law-of-large-numbers argument is related to density-dependent Markov limits~\cite{Kurtz1978} and uses conditional independence within each round. Controlled collective limits for dense open quantum systems have also been developed in Refs.~\cite{Fiorelli2023,Carollo2024}.

We take the thermodynamic limit first and then study the long-time attractors of Eq.~\eqref{eq:mfmap}. The finite-$N$ stationary laws are calculated separately in Sec.~\ref{sec:finite}; no interchange of these limits is assumed.

\subsection{Equation of state and global stability}

Write $\alpha=\alpha_\Delta\in[0,1)$ and $g(a)=\bar z_\Delta(a)$. The fixed-point condition from Eq.~\eqref{eq:mfmap} is the equation of state
\begin{equation}
 m=g(Km)=\frac{(1+\alpha)\tanh(Km)}{1+\alpha\tanh^2(Km)}.
 \label{eq:eos}
\end{equation}
Indeed,
\begin{equation}
 f(m)-m=(1-E)[1+\alpha\tanh^2(Km)]\,[g(Km)-m].
 \label{eq:signidentity}
\end{equation}
For $u=\tanh a$,
\begin{equation}
 g'(a)=\frac{(1+\alpha)(1-u^2)(1-\alpha u^2)}{(1+\alpha u^2)^2}>0.
\end{equation}
Each nonconstant positive factor on the right decreases with $u\in(0,1)$, so $g'$ strictly decreases for $a>0$. Thus $g$ is strictly concave on the positive half-line and $g'(0)=1+\alpha$.

If $K(1+\alpha)\le1$, concavity implies $g(Km)<m$ for all $m>0$. If $K(1+\alpha)>1$, the function $g(Km)-m$ has positive initial slope and is negative at $m=1$, because $g(K)<1$. Strict concavity permits precisely one positive root. Odd symmetry gives the negative root. Hence
\begin{equation}
 K_c(\Delta)=(1+\alpha_\Delta)^{-1}=\frac{1-E}{1-L}\label{eq:critical}
\end{equation}
To establish dynamical stability, put $d=E-L=\alpha(1-E)\ge0$ and $h=1-L=(1+\alpha)(1-E)$. Differentiation gives
\begin{equation}
\begin{aligned}
 f'(m)&=U(Km)+K[1-\tanh^2(Km)]\\
 &\qquad\times[h-2d\,m\tanh(Km)].
\end{aligned}
\end{equation}
Here $U\ge L>0$ and the square bracket is bounded below by $h-2d=(1-E)(1-\alpha)>0$. Thus $f$ is increasing on $[-1,1]$ for $K>0$. At a nonzero fixed point, differentiating Eq.~\eqref{eq:signidentity} and using concavity yields
\begin{equation}
 f'(m_*)-1=(1-E)(1+\alpha u_*^2)[Kg'(Km_*)-1]<0.
\end{equation}
Therefore $0<f'(m_*)<1$. For $m_0>0$, the sign in Eq.~\eqref{eq:signidentity}, together with monotonicity of $f$, gives convergence to $m_*$ above threshold and to zero below or at threshold. An increasing scalar map has no nontrivial periodic cycle. The origin above threshold has multiplier $E+hK>1$.

\subsection{Continuous onset of the ordered phase}
The expansion around the origin is
\begin{equation}
\begin{aligned}
 f(m)-m&=\epsilon m-B(K)m^3+O(m^5),\\
 \epsilon&=hK-(1-E),\\
 B(K)&=\frac{hK^3}{3}+dK^2.
\end{aligned}
\label{eq:landau}
\end{equation}
At $K_c$, the cubic restoring coefficient is
\begin{equation}
 B_c=(1-E)\frac{\alpha_\Delta+1/3}{(1+\alpha_\Delta)^2}>0.
 \label{eq:bc}
\end{equation}
Consequently,
$m_*^2=\epsilon/B_c+O(\epsilon^2)$ at fixed cadence. This confirms the supercritical pitchfork and gives the coefficient used for the fluctuation scaling.

\subsection{Ordering at fixed subcritical programmed coupling}
For the programmed Curie--Weiss law, the thermodynamic equilibrium equation is $m=\tanh(Km)$ and the zero-bias critical coupling is $1$. The measured sampler instead loses stability at Eq.~\eqref{eq:critical}. The monotonicity of $\alpha_\Delta$ gives
\begin{equation}
 K_c(0^+)=\frac{1}{1+s},\qquad K_c(\infty)=1.
\end{equation}
Thus every coupling in the open interval
\begin{equation}
 \frac{1}{1+s}<K<1
 \label{eq:window}
\end{equation}
has a unique finite critical readout interval $\Delta_c$. The measured network orders for $0<\Delta<\Delta_c$ and is disordered for $\Delta>\Delta_c$, while its programmed equilibrium law stays disordered throughout. This comparison holds $K$, $s$, $\nu$, the bias, and the measurement basis fixed, and varies only the common measurement--feedback interval.

\begin{figure}[tb]
 \includegraphics[width=\columnwidth]{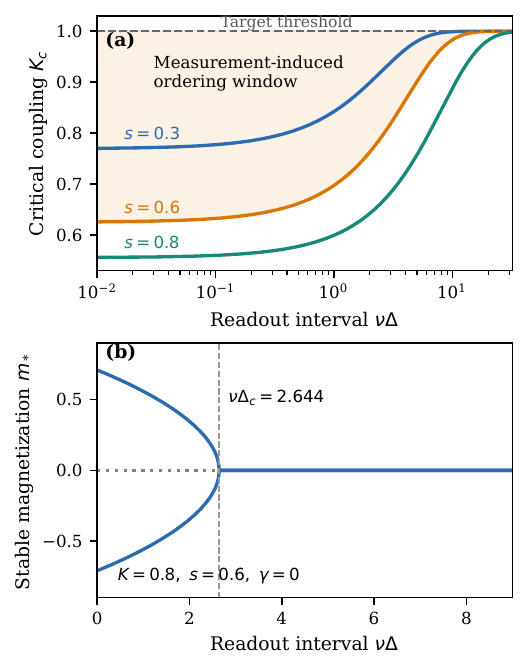}
 \caption{Ordering controlled by the readout interval, with zero environmental dephasing. (a) Exact thermodynamic threshold, Eq.~\eqref{eq:critical}. The shaded region for $s=0.6$ is ordered although the programmed Curie--Weiss law is disordered. (b) Stable solutions of Eq.~\eqref{eq:eos} at $K=0.8$, $s=0.6$. The central dotted branch is unstable. Changing the interval alone crosses the transition at $\nu\Delta_c=2.6438$.}
 \label{fig:phase}
\end{figure}

Figure~\ref{fig:phase} shows both the threshold and the stable branches. At $K=0.8$ and $s=0.6$, the transition occurs at $\nu\Delta_c=2.6438$. Below this interval a positive seed converges to $m_*>0$ and a negative seed to $-m_*$. The zero-magnetization solution is unstable below this interval. Both ordered branches are related by spin inversion, and the finite-size signature is the weight of the two magnetized sectors, as quantified in Sec.~\ref{sec:finite}.

Several controls delimit the mechanism. A full reset, $s=0$, makes the decay factors equal and removes the gain enhancement; its threshold is $K_c=1$ at every interval. At fixed $s>0$, widely separated measurements allow local relaxation to complete and recover the same thermodynamic threshold. For $K\le 1/(1+s)$, no finite interval induces a stable nonzero scalar fixed point. The relation between these thresholds and the full finite-size sampling law is discussed in Sec.~\ref{sec:finite}.

\begin{figure}[tb]
 \includegraphics[width=\columnwidth]{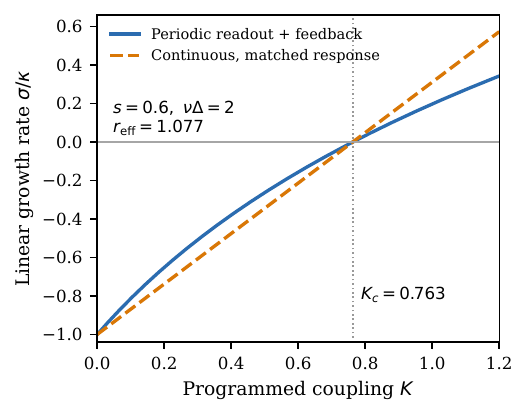}
 \caption{Linear relaxation at matched static response. At $s=0.6$ and $\nu\Delta=2$, the continuous model uses $r=r_{\rm eff}(\Delta)$ from Eq.~\eqref{eq:staticmapping}. The curves show Eq.~\eqref{eq:matchedrates} in physical time, normalized by $\kappa$. They share the critical coupling but have different relaxation rates below it and different growth rates above it. The comparison concerns infinitesimal perturbations about $m=0$, not relaxation within an ordered branch.}
 \label{fig:relaxation}
\end{figure}

Near $\Delta_c$, the linear excess gain is $\epsilon(\Delta)=(1-E)[KG_\Delta-1]$. Its derivative at the crossing is $(1-E_c)K\,\partial_\Delta G_\Delta|_c<0$. The same cubic expansion therefore describes a cadence-driven pitchfork: $m_*^2$ is proportional to $\Delta_c-\Delta$ on the ordered side. Scanning the measurement--feedback interval thus crosses the bifurcation while keeping the programmed Boltzmann distribution fixed.
The matched equation of state can be supplemented by a direct comparison of relaxation in physical time. Near $m=0$, the measured map gives $\delta m_{k+1}=[E+(1-L)K]\delta m_k$. Define $\sigma$ by $\delta m\propto e^{\sigma t}$, so negative $\sigma$ denotes decay. The two growth rates are
\begin{equation}
\begin{aligned}
 \sigma_\Delta&=\frac{1}{\Delta}\log\!\left[E+(1-L)K\right],\\
 \sigma_r&=\kappa\left[KG_\Delta-1\right],\qquad r=r_{\rm eff}(\Delta).
\end{aligned}
\label{eq:matchedrates}
\end{equation}
The second expression follows by linearizing the continuously interacting Curie--Weiss model about its unmagnetized stationary state~\cite{Zhang2026}. Both rates vanish at $K_c$, but their slopes there differ by the factor $(1-E)/(\kappa\Delta)<1$. Figure~\ref{fig:relaxation} shows this difference at fixed $s,\nu,\Delta$. It cannot be removed over the whole coupling range by a single rescaling of time, since $\sigma_\Delta$ is logarithmic in $K$ and $\sigma_r$ is linear. At $K=0$ both give $-\kappa$. This comparison quantifies a dynamical consequence of the two protocols; it includes the effect of synchronous feedback already identified in Sec.~\ref{sec:comparison}.

\section{Finite-size evidence and critical fluctuations}\label{sec:finite}
\subsection{Exact count process}

At zero bias, let $n$ be the number of positive spins and $m_n=2n/N-1$. The two possible fields, including self-field exclusion, are
\begin{equation}
 a_+=K(m_n-N^{-1}),\qquad a_-=K(m_n+N^{-1}).
\end{equation}
An initially positive spin is positive in the next round with probability $p_+=[1+U(a_+)+D(a_+)]/2$; an initially negative spin does so with probability $p_-=[1-U(a_-)+D(a_-)]/2$. Therefore
\begin{equation}
\begin{aligned}
 n'&=X_++X_-,\\
 X_+&\sim\operatorname{Bin}(n,p_+),\\
 X_-&\sim\operatorname{Bin}(N-n,p_-),
\end{aligned}
\label{eq:binomials}
\end{equation}
with independent binomials conditional on $n$. The column-stochastic matrix is
\begin{align}
 P_{n',n}&=\sum_{\ell=\max(0,n'-(N-n))}^{\min(n,n')}
 b_n(\ell;p_+)b_{N-n}(n'-\ell;p_-),\\
 b_r(j;p)&=\binom{r}{j}p^j(1-p)^{r-j}.
\end{align}
All entries are strictly positive at finite parameters. Perron--Frobenius theory gives a unique stationary vector $\pi$ with $P\pi=\pi$. Inversion symmetry gives $\pi_n=\pi_{N-n}$ exactly. Thus ordering at finite size means a bimodal distribution or a large second moment, not a nonzero stationary first moment.

The target count distribution is
\begin{equation}
 \pi_n^{\rm tar}\propto \binom{N}{n}
       \exp\!\left[\frac{K}{2N}(2n-N)^2\right].
 \label{eq:targetcount}
\end{equation}
The constant self-term has canceled from its normalization. As a sampling reference, Eq.~\eqref{eq:targetcount} is the invariant law of asynchronous heat-bath dynamics with flip rates $\nu[1-z_i\tanh a_i]/2$~\cite{Glauber1963}. Parallel conditional updates generally have a different invariant law~\cite{Peretto1984,Grinstein1985}. For example, two unbiased spins with coupling $J$ at $\Delta\to\infty$ obey $\langle z_1'z_2'\rangle=\tanh^2J\,\langle z_1z_2\rangle$, giving zero stationary correlation for finite $J$, whereas the target correlation is $\tanh J$. Thus recovery of $K_c=1$ alone does not imply recovery of the target distribution.

The stationary distribution is obtained by direct linear algebra on the exact count transition matrix, with inversion symmetry used to improve numerical conditioning; the reconstructed solution is checked against the full matrix.

\subsection{Stationary signatures of induced order}
\begin{figure}[tb]
 \includegraphics[width=\columnwidth]{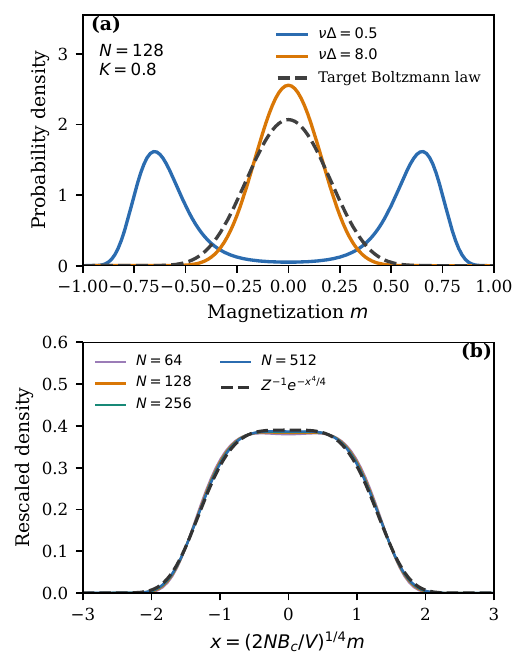}
 \caption{Exact stationary calculations for the measured finite network. (a) At $N=128$, $K=0.8$, $s=0.6$, the output is bimodal for $\nu\Delta=0.5$ and unimodal for $\nu\Delta=8$. The dashed curve is the programmed law, Eq.~\eqref{eq:target}. (b) At $K=K_c(0.5/\nu)=0.661927$, rescaled distributions approach the quartic critical form. Here $V=1-E^2$ and $B_c$ is given by Eq.~\eqref{eq:bc}. All curves are theoretical; no trajectory sampling is used.}
 \label{fig:statistics}
\end{figure}

Figure~\ref{fig:statistics}(a) compares two readout intervals at identical programmed weights. For $N=128$, $K=0.8$, $s=0.6$, and $\nu\Delta=0.5$, the stationary distribution peaks at $|m|=0.65625$, close to the thermodynamic attractor $m_*=0.63904$. Increasing the interval to $8/\nu$ restores a central peak. The second moments are $0.37064$ and $0.02323$, respectively. The probabilities $\Pr(|M_N|>0.3)$ are $0.9487877$ and $0.0434795$. The dashed target distribution stays centered at zero in both comparisons.

These distributions are stationary solutions of the finite count process, so their bimodality is not a transient of the initial condition. Switching between the two signs enforces a zero stationary mean at finite $N$; the second moment and the weight of the magnetized sectors resolve the ordering. The target law remains unimodal, making the discrepancy a change in the shape of the sampled distribution as well as in its width.
\subsection{Critical window and scaling}

At the origin, the conditional variance is
\begin{equation}
 \Var(M_N'|M_N=0)=\frac{1-E^2}{N}+O(N^{-2})=\frac{V}{N}+O(N^{-2}).
\end{equation}
Near criticality, the conditional variance changes by $O(m^2/N)$ and the drift has the expansion in Eq.~\eqref{eq:landau}. On the scale $m=O(N^{-1/4})$, $\epsilon=O(N^{-1/2})$, the process evolves slowly over $O(N^{1/2})$ rounds at fixed cadence. A leading Kramers--Moyal expansion gives
\begin{equation}
 \partial_k p=-\partial_m[(\epsilon m-B_cm^3)p]
                   +\frac{V}{2N}\partial_m^2p .
 \label{eq:fp}
\end{equation}
This is an asymptotic description of the critical window; away from it, the exact transition matrix is the appropriate calculation. The zero-current stationary solution is
\begin{equation}
 p(m)\propto\exp\!\left[\frac{N\epsilon}{V}m^2-
                         \frac{NB_c}{2V}m^4\right].
\end{equation}
Define
\begin{equation}
 x=\left(\frac{2NB_c}{V}\right)^{1/4}m,\qquad
 t=\epsilon\sqrt{\frac{2N}{B_cV}}.
 \label{eq:scaling}
\end{equation}
The scaling form is $p(x)\propto e^{tx^2/2-x^4/4}$. At $t=0$, the normalization is $Z=\Gamma(1/4)/\sqrt2$, and
\begin{align}
 \langle x^2\rangle&=\frac{2\Gamma(3/4)}{\Gamma(1/4)}=0.67597824\ldots,\\
 U_4&=1-\frac{\langle m^4\rangle}{3\langle m^2\rangle^2}\nonumber\\
 &=1-\frac{\Gamma(1/4)^2}{12\Gamma(3/4)^2}=0.27052013\ldots .
\end{align}
The quartic distribution is standard mean-field critical behavior~\cite{Ellis1978}; the associated slow critical dynamics is consistent with the familiar role of the critical window in Curie--Weiss sampling~\cite{Ding2009}. The cadence-dependent coefficients $B_c$ and $V$ are supplied by the present exact measured kernel. Figure~\ref{fig:statistics}(b) and Table~\ref{tab:critical} compare the resulting parameter-free prediction with the exact finite-size distributions. Both the rescaled second moment and the Binder cumulant approach their critical values as $N$ increases. The scale factor contains information beyond the static equation of state:
\begin{equation}
 \frac{B_c}{V}=\frac{\alpha_\Delta+1/3}{(1+\alpha_\Delta)^2(1+E)}.
 \label{eq:fluctuationscale}
\end{equation}
Thus the static response parameter $\alpha_\Delta$ alone does not specify the fluctuation amplitude across protocols with different $s$ and $\Delta$.

\begin{table}[ht]
\caption{Exact finite-size checks at $s=0.6$, $\nu\Delta=0.5$, and $K=K_c=0.661926637$. No fitted scale factors are used.}\label{tab:critical}
\begin{ruledtabular}
\begin{tabular}{rcc}
$N$ & $\langle x^2\rangle$ & $U_4$\\
64 & 0.6655230 & 0.3040638\\
128 & 0.6683586 & 0.2934901\\
256 & 0.6704721 & 0.2863761\\
512 & 0.6720236 & 0.2815346\\
Critical prediction & 0.6759782 & 0.2705201
\end{tabular}
\end{ruledtabular}
\end{table}

\section{Extension to dense population networks}
For dense networks with a finite number of populations, the record kernel gives a closed limiting map and a linear instability criterion. The local factors $E,L$ encode the measurement interval; the population coupling matrix specifies the network.

\subsection{Population dynamics}

Keep zero bias and common local parameters, and partition the spins into a fixed number $B$ of populations, with $N_a/N\to f_a>0$ and $\sum_a f_a=1$. Use symmetric weights $W_{ab}/N$ between populations, omit the self coupling, and let $m_a=N_a^{-1}\sum_{i\in a}z_i$. The field in population $a$ concentrates at $h_a=\sum_b W_{ab}f_bm_b=(C\bm m)_a$. The same conditional-variance argument, now with bound $1/N_a$, proves the finite-time limit
\begin{equation}
 m_a'=U(h_a)m_a+D(h_a),\qquad C_{ab}=W_{ab}f_b.
 \label{eq:populationmap}
\end{equation}
Self-field omission contributes $O(N^{-1})$. The concentration argument uses dense couplings; at finite degree, the input field generally retains order-one fluctuations.

\subsection{Linear threshold and its domain}

Assume that $W$ is real symmetric, nonnegative, nonzero and irreducible. If $F=\diag(f_1,\ldots,f_B)$, then $C=WF$ is similar to the symmetric matrix $F^{1/2}WF^{1/2}$ and has real eigenvalues. Its largest eigenvalue $\lambda>0$ is simple and has a strictly positive eigenvector. Since $U'(0)=0$ and $D'(0)=h=1-L$, the Jacobian of Eq.~\eqref{eq:populationmap} is
\begin{equation}
 J_0=E\Id+hC.
\end{equation}
The multiplier of mode $j$ is $\Lambda_j=E+h\lambda_j$. The Perron bound gives $|\lambda_j|\le\lambda$. Before $h\lambda$ reaches $1-E$, all multipliers obey $\Lambda_j<1$ and $\Lambda_j\ge E-h\lambda>2E-1>-1$. Thus the first loss of stability is the Perron multiplier reaching $+1$:
\begin{equation}
 \lambda\frac{h}{1-E}=\lambda G_\Delta=1.
 \label{eq:spectralthreshold}
\end{equation}
For a symmetric signed $W$, the necessary and sufficient linear condition is instead
\begin{equation}
 -\frac{1+E}{h}<\lambda_j<\frac{1-E}{h}\quad\hbox{for every }j.
\end{equation}
A negative eigenvalue can then reach multiplier $-1$. Directed matrices can have complex eigenvalues and require a separate unit-disk analysis. Nonuniform readout parameters give a Jacobian $\diag(E_a)+\diag(1-L_a)C$, so a single scalar gain generally does not suffice. Equation~\eqref{eq:spectralthreshold} identifies the first linear instability; the nonlinear attractors of general population networks require a separate analysis.

\section{Discussion and conclusions}
A network of partially resetting qubits can change from disordered to ordered output as its measurement--feedback interval is reduced, while the programmed weights and local dissipative rule are held fixed. Projective readout removes the transverse component that develops during relaxation. For $s>0$, this increases the local measured gain and shifts the Curie--Weiss bifurcation below the critical coupling of the reference Boltzmann law. The limiting map has two stable nonzero branches, and exact finite-size calculations show the corresponding symmetric bimodal distributions and mean-field critical scaling.

The relation to continuous dephasing is precise at the level of the clamped response and the Curie--Weiss equation of state. The record kernel supplies the additional dynamical information: matched static responses have different relaxation rates, and the finite-size critical width depends on the decay factor $E$ as well as on the response parameter. Holding feedback fields fixed also changes multispin transitions relative to continuous classical updates. These schedule effects accompany the local quantum response enhancement and are relevant when predicting sampling statistics from programmed weights.

For dense ferromagnetic population networks, the same calculation gives the linear threshold $\lambda G_\Delta=1$. The analysis assumes ideal projective readout, instantaneous feedback, uniform local parameters, and dense couplings. It establishes the attractors after taking the thermodynamic limit and checks stationary statistics at finite sizes; a general stationary-limit theorem is outside its scope. Finite readout duration, feedback delay, and input fluctuations on sparse graphs remain to be included before applying these thresholds to a specific device.

\begin{acknowledgments}
This work is supported by the Key Research and Development Support Program of the Chengdu Municipal Science and Technology Bureau through the project ``Research, Development, and Application Demonstration Platform for an Independently Controllable Room-Temperature Quantum--Supercomputing Integrated System'' (Grant No. 2025-XT00-00012-GX). KLC acknowledges funding from the National Research Foundation, Singapore and the Ministry of Education, Singapore.
\end{acknowledgments}
\bibliographystyle{apsrev4-2}
\bibliography{references}
\end{document}